# Inverted structure perovskite solar cells: A theoretical study

[a,b]Anurag Sahu, [a]Ambesh Dixit

[a] Department of Physics and Centre for Solar Energy , Indian Institute of Technology Jodhpur, Rajasthan, 342037, India

[b] Centre for System Science , Indian Institute of Technology Jodhpur, Rajasthan, 342037, India

*Corresponding author. Tel.: +91-2912449045

E-mail address: ambesh@iitj.ac.in (Ambesh Dixit)

**Abstract:** We analysed perovskite $CH_3NH_3PbI_{3-x}Cl_x$ inverted planner structure solar cell with nickel oxide (NiO) and spiro-MeOTAD as hole conductors. This structure is free from electron transport layer. The thickness is optimized for NiO and spiro-MeOTAD hole conducting materials and the devices do not exhibit any significant variation for both hole transport materials. The back metal contact work function is varied for NiO hole conductor and observed that Ni and Co metals may be suitable back contacts for efficient carrier dynamics. The solar photovoltaic performance showed a linear decrease in efficiency with increasing temperature. The electron affinity and band gap of transparent conducting oxide and NiO layers are varied to understand their impact on conduction and valence band offsets. A range of suitable band gap and electron affinity values are found essential for efficient device performance.

## 1. Introduction:

Perovskite solar cells (PSCs) are getting focus in photovoltaic community since their introduction in 2009 [1]. PSCs efficiency evolved from 3.81% in 2009 to 19.7% efficiency over the years rapidly [2]. Initial PSCs are reported in sensitized solar cells geometries, [1], Later solid hole conductors e.g. 2,2′,7,7′-tetrakis (N,N-di-p-methoxyphenyl-amine) 9,9′-spirobifluorene (spiro-MeOTAD) are used and fairly good efficiencies are observed [3],[4],[5]. Earlier PSCs were reported in semiconducting/insulating mesoporous electrode configuration, similar to that of sensitized excitonic solar cells, which are used as scaffold only to deposit perovskite in later device structures [6],[4]. This demonstrated that perovskite solar cells are different from the conventional sensitized excitonic solar cells, and electron injection to electron transport material is not mandatory in PSCs. Later, planner junction perovskites are also investigated and advanced to fairly high efficiencies [7],[8],[9]. PSCs are also explored in hole transport layer free structure [10], [11] and electron transport layer free structure configurations [12],[13],[14]. Inorganic – organic halide perovskite absorbers showed good carrier diffusion length in $\mu m$ range together with large absorption coefficient ~ $10^5$ $cm^{-1}$ across the entire solar spectrum. These electronic and optical properties make such inorganic-organic halide perovskites efficient absorber candidates for rapidly evolving photovoltaic field [15],[16].

One dimensional device simulation is used intensively investigating thin film solar cells such as $Cu(In,Ga)Se_2$ (CIGS) [17],[18], CdTe [19],[20], $CuSnI_3$ (CIS) [21], and $Cu_2ZnSn(S,Se)_4$ (CZTS) [22] over the years for optimization of thickness of layered structures and to observe the effect of device configuration, materials parameter on their photovoltaic performance. There are also few reports on device simulation for planner configurations of perovskite solar cells in recent years [23],[24],[25],[26],[27],[28]. Considering the configurational similarity of organic-inorganic hybrid perovskite solar cells with inorganic material based solar cells like CZTS, CIGS; and low binding energy exciton (Wannier type exciton) [29] in perovskites, device simulation is successfully employed with 1D simulator like solar cell capacitance simulator (SCAPS) [30] and analysis of microelectronics and photonics structure (AMPS) [23]. A reasonable agreement is observed between simulated and experimental results for device performance. The planner PSCs, which are usually simulated, consist of transparent conducting oxide (TCO)/planner $TiO_2$/perovskite/(HTM) hole transport material/metal contact, as shown schematically in Fig 1(a). Further, ETM free planner perovskite devices, as shown in Fig 1(b), are also simulated, where electron transport layer like $TiO_2$ is not considered [28].

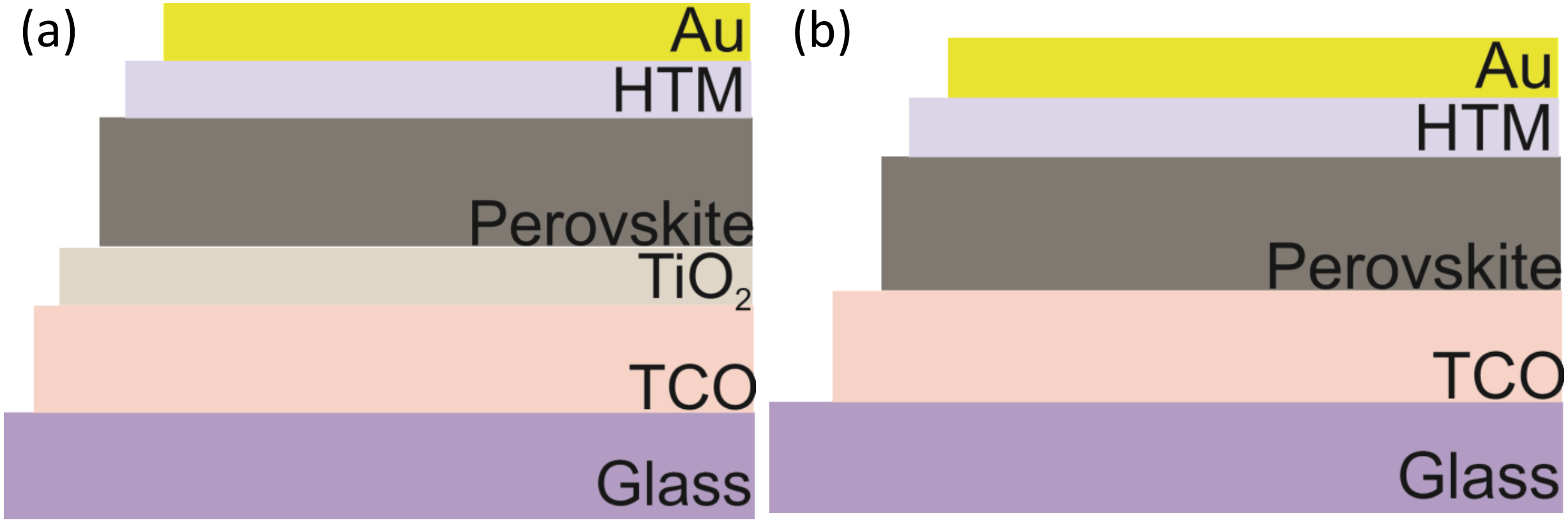

**Figure 1:** Schematic diagrams for planner (a) and ETM free planner (b) perovskite device structures, showing different layer stacking on glass substrates.

We simulated inverted PSC structures which are considered as ETL free PSC devices. The inverted device structure can be realized by depositing hole conductor like spiro –MeOTAD or NiO layer on metal substrates like Ni, followed by transparent conducting oxide (TCO) layer deposition for the efficient collection of photo generated carriers. A metal contact can be deposited on TCO. This is similar to an ETM free device structure reported earlier and the schematic device structure with different layers is shown in Fig 2. The proposed inverted planner perovskite device structure is simulated using 1D simulator SCAPS [30]. We optimized device structures for both NiO and spiro-MeOTAD hole conductor material for different perovskite absorber thicknesses. We also simulated the effect of back metal substrate work function for different hole conductors on photovoltaic performance. This will assist in finding potential metal substrates for deposition of NiO or spiro-MeOTAD hole conductor material. We studied effect of conduction band offset (CBO) and valence band offset (VBO) for inverted planner PSCs structure. In this case, CBO is evaluated at TCO/perovskite interface and VBO is evaluated at perovskite/HTM interface. This inverted planner perovskite device structure seems very promising, where the top TCO layer will cover perovskite absorber and may act as isolation/protecting layer to environmental conditions for better stability. However, the experimental realization of these structures is yet to carried out.

**2. Device Simulation parameter:**

Inverted PSCs considered for simulation is shown schematically in Fig 2, consisting of back metal contact/TCO/Perovskite/HTM/metal substrate. All materials' parameters used for different layers are listed in Table 1. TCO and absorber material parameters are based on FTO ($SnO_2$:F) and $CH_3NH_3PbI_{3-x}Cl_x$ perovskite. Initial perovskite absorber, TCO, and HTM thicknesses are considered close to the reported experimental values [5]. Absorber defects are considered neutral and defect density is considered about $1.3 \times 10^{14}$ cm$^{-3}$ to provide diffusion length of 1µm in perovskite absorber, close to the reported experimental value [5]. Thermal velocities for electrons and holes are taken to be $10^7$ cm s$^{-1}$ for all materials. Defect centres are considered in the mid band by selecting their positions below the conduction level equal to half of band gap and a Gaussian distribution is considered with characteristic energy 0.1eV [24],[31]. Neutral defect type is utilized while electron and hole capture cross section of about $2 \times 10^{-14}$ cm$^2$ is considered for simulation. Defect density is considered about $10^{15}$ cm$^{-3}$ for TCO and hole conductor. A pre-factor for $10^5$ cm$^{-1}$ absorption coefficient is used for absorber layer as reported earlier [24]. Optical reflectance at surface and interfaces are not considered in this simulation. Flat band conditions are opted for top contact at TCO and bottom metal substrate below NiO layer.

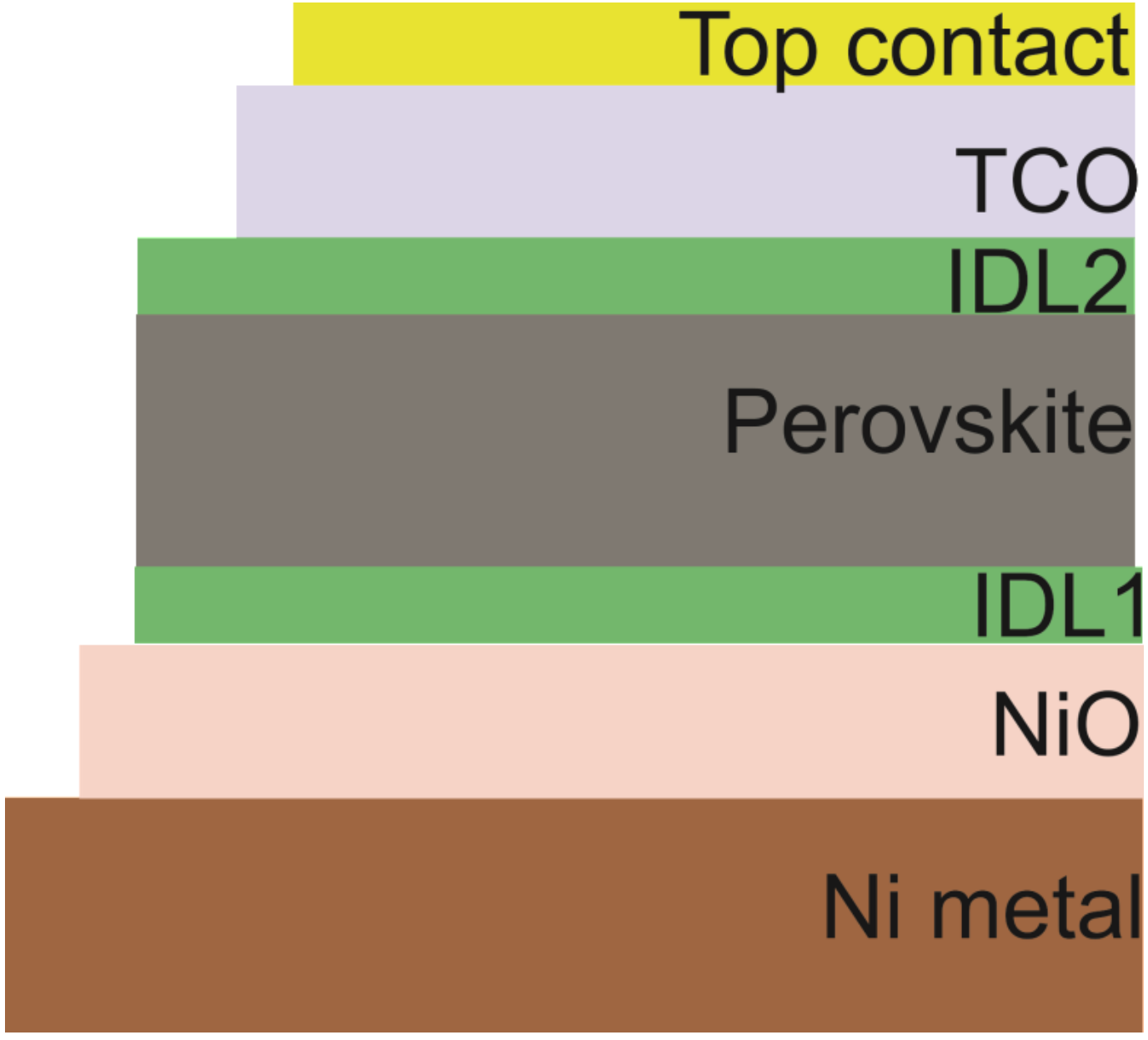

**Figure 2**: Schematic diagram of inverted planner perovskite solar cell device considered for simulation.

Table: 1 Material parameter for each layer in inverted planner perovskite solar cell for simulation of initial structure.

| Material Property | TCO [25] | Perovskite [32] | NiO [32],[31] | IDL1 | IDL2 | Spiro-MeOTAD [32] |
|---|---|---|---|---|---|---|
| Thickness (nm) | 500 | 450 | 450 | 10 | 10 | 450 |
| Dielectric Permittivity (ε) | 9 | 10 [33] | 10.7 [34] | 10 | 10 | 3 |
| Electron Mobility $\mu_n$ ($cm^2 V^{-1}s^{-1}$) | 20 | 10 [35] | 12[36] | 1E+1 | 1E+1 | 1E-4 |
| Hole Mobility $\mu_p$ ($cm^2 V^{-1}s^{-1}$) | 10 | 10 [35] | 2.8 [37] | 1E+1 | 1E+1 | 2E-4 |
| Acceptor | 0 | 1E+9 | 1E+18 | 1E+9 | 1E+9 | 1E+18 |
| Concentration $N_A$ cm$^{-3}$) | | | | | | |
| Donor Concentration $N_D$ (cm$^{-3}$) | 2E+19 | 1E+9 | 0 | 1E+9 | 1E+9 | 0 |
| Bandgap $E_g$ (eV) | 3.5 | 1.55 [38] | 3.6 [39] | 1.55 | 1.55 | 3.06 |
| CB DOS (cm$^{-3}$) | 2.8E+18 | 2.75E+18 [40] | 2.8E+18 [41] | 2.75E+18 | 2.75E+18 | 2.8E+19 |
| VB DOS (cm$^{-3}$) | 1.8E+19 | 3.9E+18 [40] | 1E+19 [41] | 3.9E+18 | 3.9E+18 | 1E+19 |
| Affinity χ (eV) | 4 | 3.9 [29] | 1.8 [39] | 3.9 | 3.9 | 2.05[42] |

The device structure is simulated using 1D-SCPAS which primarily solves following coupled equations for bulk layers.

$$\frac{\partial\left(\varepsilon\frac{\partial\varphi}{\partial x}\right)}{\partial x} = -q\left(p - n + N_D^+ - N_A^- + \frac{\rho_{def}}{q}\right) \quad (1)$$

$$-\frac{\partial J_n}{\partial x} - U_n + G = \frac{\partial n}{\partial t} \quad (2)$$

$$-\frac{\partial J_p}{\partial x} - U_p + G = \frac{\partial p}{\partial t} \quad (3)$$

These equations are solved along with following equations 4 and 5. Here $\varepsilon$ is dielectric constant, $\varphi$ is electric potential, q is magnitude of electronic charge, p is hole concentration, n is electron concentration, $N_D^+$ is donor density concentration, $N_A^-$ is acceptor concentration, $\rho_{def}$ is charge density of layer, $J_n$ is electron current density, $U_n$ is electron recombination, G is generation rate, $\frac{\partial n}{\partial t}$ is rate of change of electron concentration, $J_p$ is hole current density, $U_p$ is hole recombination and $\frac{\partial p}{\partial t}$ is rate of change of hole concentration.

$$J_n = -\frac{\mu_n n}{q}\frac{\partial E_{Fn}}{\partial x} \tag{4}$$

$$J_p = \frac{\mu_p p}{q}\frac{\partial E_{Fp}}{\partial x} \tag{5}$$

Here $\mu_n$ is electron mobility, $E_{Fn}$ is electron Fermi level, $\mu_p$ is hole mobility and $E_{Fp}$ is hole Fermi level. These equations are solved numerically by employing a denser mesh at interface and coarse mesh in layer with appropriate boundary conditions at interfaces and contacts. However, meshing is adoptive during calculation. Gummel iteration scheme along with Newton-Raphson substeps is utilized in SCAPS for solving these coupled equations.

Interface defect layers IDL1 and IDL2 are inserted at respective interface to consider interface carrier recombination at absorber/TCO and absorber/HTM interfaces. All materials properties of defect layers are considered similar to that of perovskite layer except defect density of interface layer is set at $10^{17}$ $cm^{-3}$ to model defects at interface [26]. The initial simulation is carried out with the parameters, summarized in Table 1 for NiO hole conductor and results are shown in Fig 3.

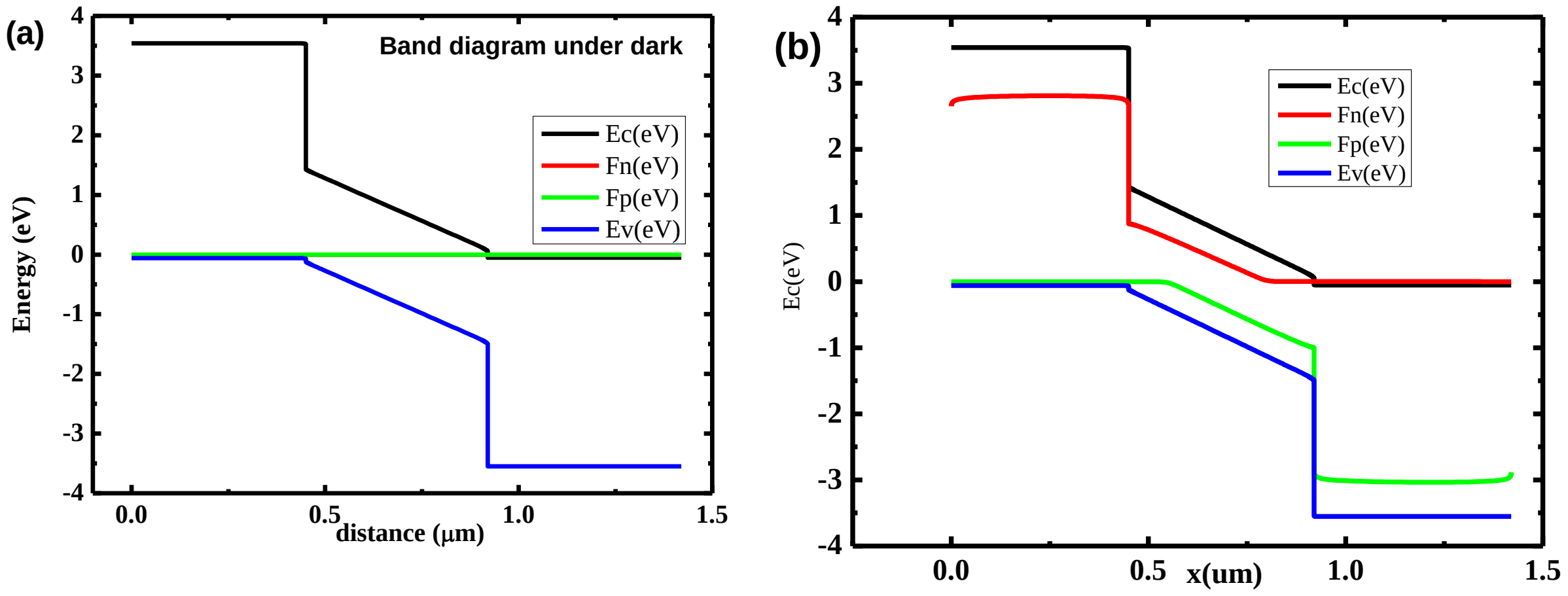

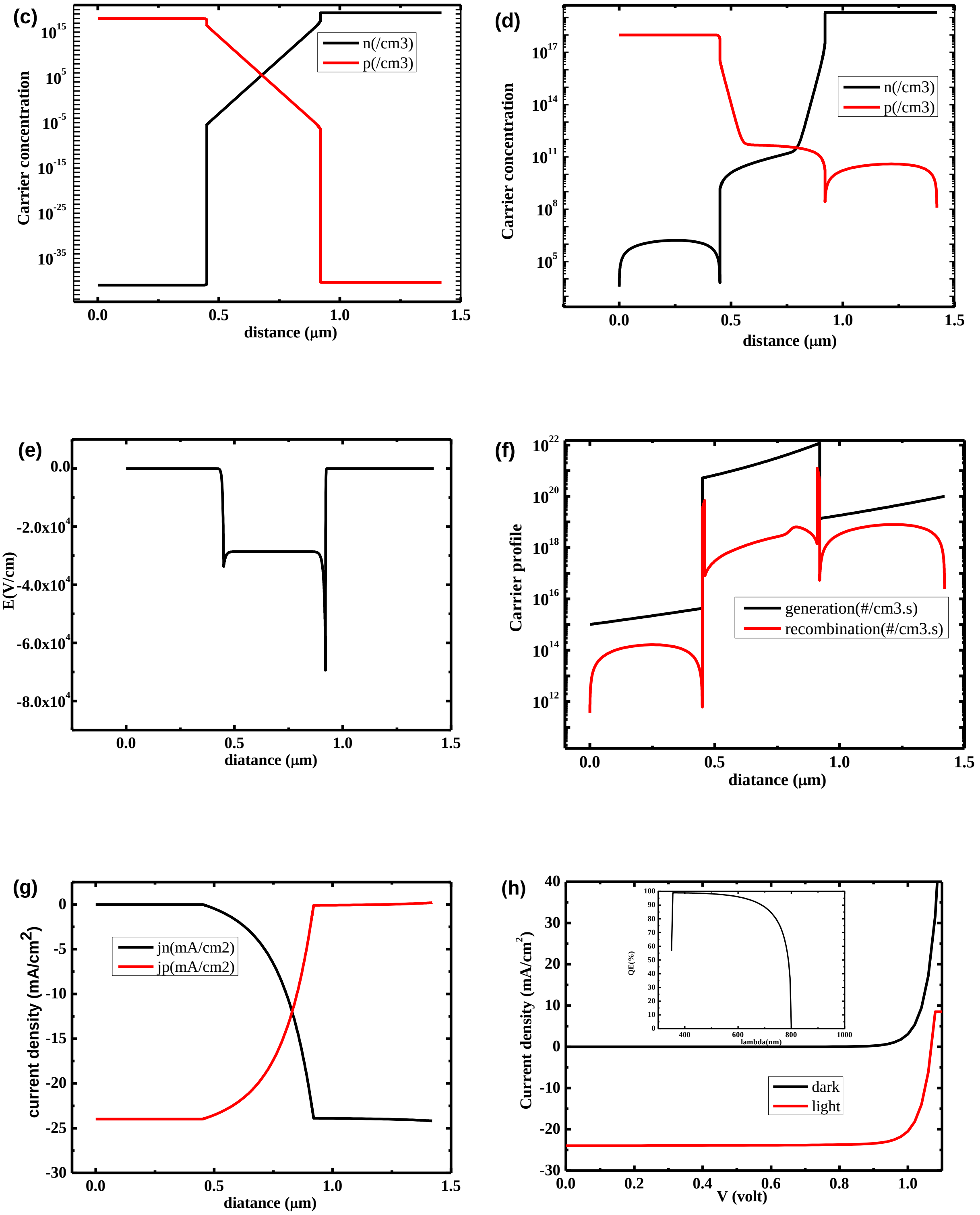

**Figure 3**: Simulation results for inverted PSC structure under light and dark (a) energy level diagram of solar cell under dark, (b) energy level diagram of solar cell under illumination, (c) carrier concentration in solar cell under dark, (d) carrier concentration in solar cell under illumination, (e)

electric field intensity across solar cell, (f) generation and recombination of photo generated carriers across device, (g) electron and hole current density across device under illumination, and (h) Current-voltage characteristic of simulated device under AM 1.5 spectrum and quantum efficiency of device (in inset).

Figures 3 (a) and 3 (b) are showing band bending for simulated device under dark and light conditions, respectively. Figure 3 (a) explains that Fermi levels for electrons and holes are pinned in conjunction with hole Fermi-level lying in NiO valence band and electron Fermi level lying in FTO conduction band under equilibrium. Figure 3(b) shows unpinning of Fermi level under illumination. Electron Fermi level resides in FTO conduction band while it is close to NiO conduction band. Similarly, hole Fermi level resides in NiO valence band and near to FTO valence band. This suggests that there is a significant concentration of minority carriers in FTO and NiO sides.

Figures 3 (c) and 3 (d) are showing carrier concentration under dark and light. Figure 3 (c) explains that under dark, there are negligible electrons and holes in both NiO hole conductor and TCO layers. The carrier concentration varies linearly in depletion region of absorber. Figure 3 (d) suggests that hole and electron concentrations in NiO and TCO layers are equal to intrinsic carrier concentrations in NiO and TCO, respectively. A significant change in minority carriers is observed under illumination in both NiO and TCO layers. The minority carrier concentration is high, close to TCO/absorber and absorber/HTM interfaces and is constant across NiO and TCO, resulting to the lower concentration at contacts. Hole concentration is the highest near NiO and absorber interface, which reduces as one moves toward TCO side in the absorber. However, hole concentration decreases slowly under illumination due to the hole generation in absorber. A similar trend is observed for variation of electron concentration in device under illumination. Further, a sharp reduction in carriers near defect layers can be observed due to the higher defect density in respective interface layers.

Electric field intensity variation across the device is plotted in Fig 3 (e). Higher field intensity is found at interfaces, whereas steady built in field is found deep inside absorber. There is no effect on electric field intensity against illumination. Generation and recombination profile for carriers is plotted in Fig 3 (f). The generation of carrier decreases linearly from TCO side (light incident side) to NiO side in absorber and a similar trend is observed in TCO and NiO layers. Higher carrier generation is observed in TCO side due to a slightly lower band gap (3.5eV) as compared to NiO (3.6eV) and also due to the illumination from TCO side. Recombination is also showing linear reduction from illumination side as recombination also depends on carrier concentration, which is higher towards illumination side in absorber and lower on other side.

The onset of cliff at interface is attributed to the enhanced interface recombination. Further, recombination rates are nearly constant in TCO and NiO layer, showing sharp reduction at contacts, suggesting good selectivity because of the flat band conditions opted for simulation of contacts. Current density profile is plotted in Fig 3 (h), with inset showing quantum efficiency. The short circuit current density $J_{sc}$, open circuit potential $V_{oc}$ and efficiency values are 24mA/cm$^2$, 1.07 Volt and 21.66 % for the present device structure with very small (nano ampere) dark current. The calculated efficiency value is close to the simulated photovoltaic efficiency for planner PSC structure reported for spiro-MeOTAD hole conductor showing validity of simulation conditions adopted in the present work [28]. We further extended the simulation studies to optimize the individual materials layer in inverted planner PSCs to understand their impact on device performance.

## 3. Results and Discussion:

### 3.1 Device structure optimization:

The initial inverted PSC, Fig 2, is considered for present investigations and simulations are carried out for a set of parameters, listed in Table 1. The results are discussed in earlier section. Further, to understand the impact of each layer thickness on device performance, we varied the thickness of transparent conducting oxide (TCO) layer and optimized for a thickness with the optimal efficiency. The absorber and hole conductor layers thicknesses are varied simultaneously considering the optimal TCO thickness for finding out the optimum combination of absorber and hole transport layer thicknesses. We considered TCO thickness from 2µm to 100nm thickness and restricted the lowest thickness at 100 nm for any device layer during optimization, considering the practical deposition limitations of such thin layers and possibility of higher defect density for lower layer thicknesses. We observed that the photovoltaic efficiency of the devices increases with reducing the TCO thickness, as shown in Fig 4(a) for NiO hole conductor. However, the calculated changes are not significant and the efficiency variation is within 0.3% for 1.8 µm variation in TCO thickness. Considering the non-significant effect of TCO layer thickness, we considered 100nm TCO layer thickness while varying absorber and NiO hole conductor thicknesses simultaneously. Absorber thickness is varied from 2µm to 100 nm with 100 nm thickness step and similar variation is also considered for NiO hole conductor simultaneously. The variation in efficiency is plotted in Fig 4 (b). We can see clearly that there is no significant effect on efficiency for NiO hole conductor thickness, whereas there is a considerable impact of perovskite absorber thickness in the device performance. The photovoltaic efficiency is about 21.15% for 300nm absorber thickness and is increasing with absorber thickness with the maximum value for 600nm to 800 nm absorber thicknesses. This further starts decreasing with absorber thickness. Further, optimization is carried out by varying absorber thickness in 600nm to 800 nm range with 10nm step size and NiO thickness in 1000 nm to 100 nm thickness range with 100nm step size.

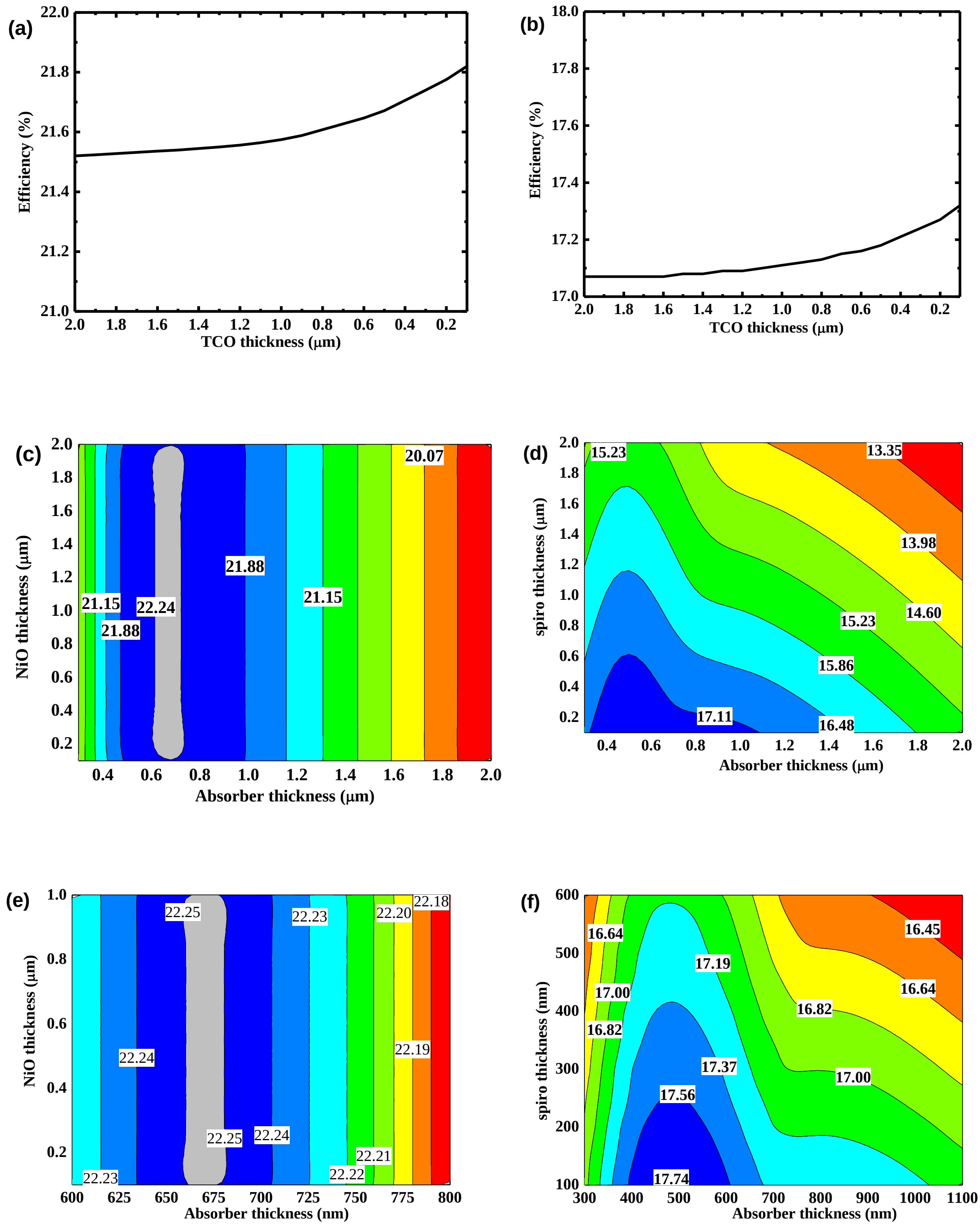

**Figure 4**: (a) Efficiency variation with TCO thickness with NiO hole conductor, (b) efficiency variation with TCO thickness with spiro-MeOTAD hole conductor, (c) efficiency variation with simultaneous change in thickness of NiO and absorber, (d) efficiency variation with simultaneous change in thickness of spiro-MeOTAD and absorber (e) efficiency variation with simultaneous change in absorber thickness in 600nm-800nm range and NiO thickness in 100nm-1000nm range, and (f) efficiency variation with simultaneous change in absorber thickness in 300nm-1100nm range and NiO thickness in 100nm-600nm range.

We observed that the highest efficiency corresponds to 650 nm perovskite absorber thickness, while the effect of NiO thickness is not significant on photovoltaic efficiency under standard material parameter, Table 1. The optimized layer thicknesses are summarized in Table 2 for inverted PSC in conjunction with NiO hole conductor.

A similar optimization process is carried out for spiro-MeOTAD hole conductor and the variation of efficiency with TCO thickness variation is plotted in Fig 4(b). Here, we observed a small increase in efficiency with decreasing TCO layer thickness and a thickness of about 100nm is selected for further calculations. The perovskite absorber and spiro-MeOTAD hole conductor thicknesses are varied in 2μm to 100 nm range simultaneously and results are plotted in Fig 4 (d). The absorber thickness in 300 nm to 1100 nm range and hole conductor thickness in range 100nm to 600nm thickness range resulted about 17% photovoltaic efficiency. Further, absorber thickness is varied in 300nm to 1100 nm range with 10nm step size in conjunction with variation in spiro-MeOTAD thickness from 600 nm to 100 nm with 100 nm step size. This suggests that 470nm perovskite absorber thickness is the optimal thickness for achieving the highest efficiency, while decrease in spiro-MeOTAD hole conductor thickness resulted in enhanced efficiency. Optimized thickness of inverted PSCs with NiO and spiro-MeOTAD hole conductors are summarized in Table 2.

**Table 2**: Optimized thickness for inverted planner perovskite solar cell with NiO and spiro-MeOTAD hole conductor.

| Absorber thickness | HTM material | HTM thickness | TCO thickness | Performance parameter | | | |
|---|---|---|---|---|---|---|---|
| | | | | $J_{SC}$ (mA/cm$^2$) | $V_{OC}$ (volt) | FF | Efficiency |
| 650nm | NiO | 100nm | 100nm | 25.15 | 1.06 | 83.36 | 22.25% |
| 470nm | Spiro-MeOTAD | 100nm | 100nm | 24.26 | 1.08 | 67.85 | 17.74% |

Additionally, we observed that the fill factor for inverted planner perovskite solar cell is relatively poor with spiro-MeOTAD hole conductor. This poor fill factor is attributed to different electron affinity values of spiro-MeOTAD hole conductor. The electron affinity of spiro-MeOTAD is considered about 2.05 eV, as listed in Table 1, for present study. However, a different value of about 2.45 eV is considered as electron affinity of spiro-MeOTAD hole conductor and is reported to be variable [26].

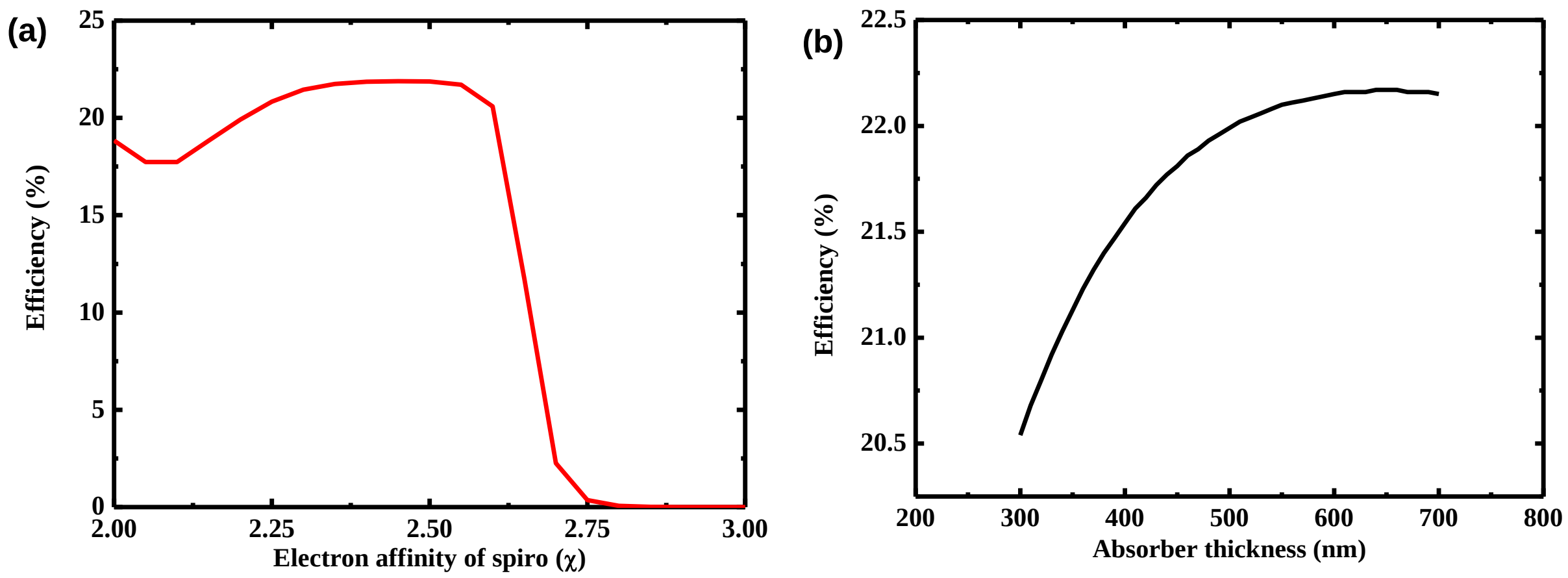

**Figure 5:** (a) Effect of electron affinity of spiro-MeOTAD over phot-voltaic performance of inverted planner perovskite solar cell and (b) Effect of absorber thickness over photovoltaic performance of inverted planner perovskite device.

Thus, electron affinity of spiro-MeOTAD hole conductor is varied to understand the effect of electron affinity on photovoltaic performance. We find significant impact of electron affinity on efficiency and observed that a value of 2.45 eV gives the maximum efficiency under considered spiro-MeOTAD hole conductor material parameters. Thus, inverted PSC with 2.45 eV electron affinity for spiro-MeOTAD hole conductor is used for further investigations. As observed that reducing spiro-MeOTAD thickness resulted into enhanced device performance, we considered 100 nm thick spiro-MeOTAD hole conductor layer for simulating devices. The perovskite absorber layer thickness is varied from 300 nm to 700 nm with a 10 nm step size. The calculated efficiency as a function of absorber thickness is plotted in Fig 5 (b). We find that device efficiency increases with increasing the absorber thickness and showed saturation trend around 22.10% for 550 nm absorber thickness with the maximum efficiency for 640 nm. This starts decreasing with further increasing the absorber thickness. The different layer optimized thicknesses are summarized in Table 3 for inverted planner perovskite device with NiO and spiro-MeOTAD hole conductors.

**Table 3:** Summary of final optimized inverted planner perovskite device with NiO and spiro-MeOATD hole conductors.

| Absorber thickness | HTM material | HTM thickness | TCO thickness | Performance parameter | | | |
|---|---|---|---|---|---|---|---|
| | | | | $J_{SC}$ (mA/cm$^2$) | $V_{OC}$ (volt) | FF | Efficiecny |
| 650nm | NiO | 100nm | 100nm | 25.15 | 1.06 | 83.36 | 22.25% |
| 640nm | Spiro-MeOTAD | 100nm | 100nm | 25.130 | 1.07 | 82.40 | 22.17% |

The photovoltaic response of optimized inverted planner perovskite device with NiO and spiro-MeOTAD hole conductors is summarized in Fig 6. Figs 6 (a) and 6 (b) show current density–voltage characteristic under dark and illumination with respective quantum efficiency for optimized devices under illumination. Figure 6(b) shows quantum efficiency for both hole conductors, considered for inverted planner perovskite solar cell. We can see that quantum efficiency starts increasing from 800nm, near the absorber band gap and attained the maximum value of 98 at 660 nm, which is constant till 355 nm and start decreasing at 355 nm due to the band gap absorption for TCO and hole conductors. Thus, inverted planner devices perform nearly identical for both hole conductors and thus, NiO can be considered as a suitable choice for such PSCs and also as a good alternative to very costly spiro-MeOTAD hole conductor.

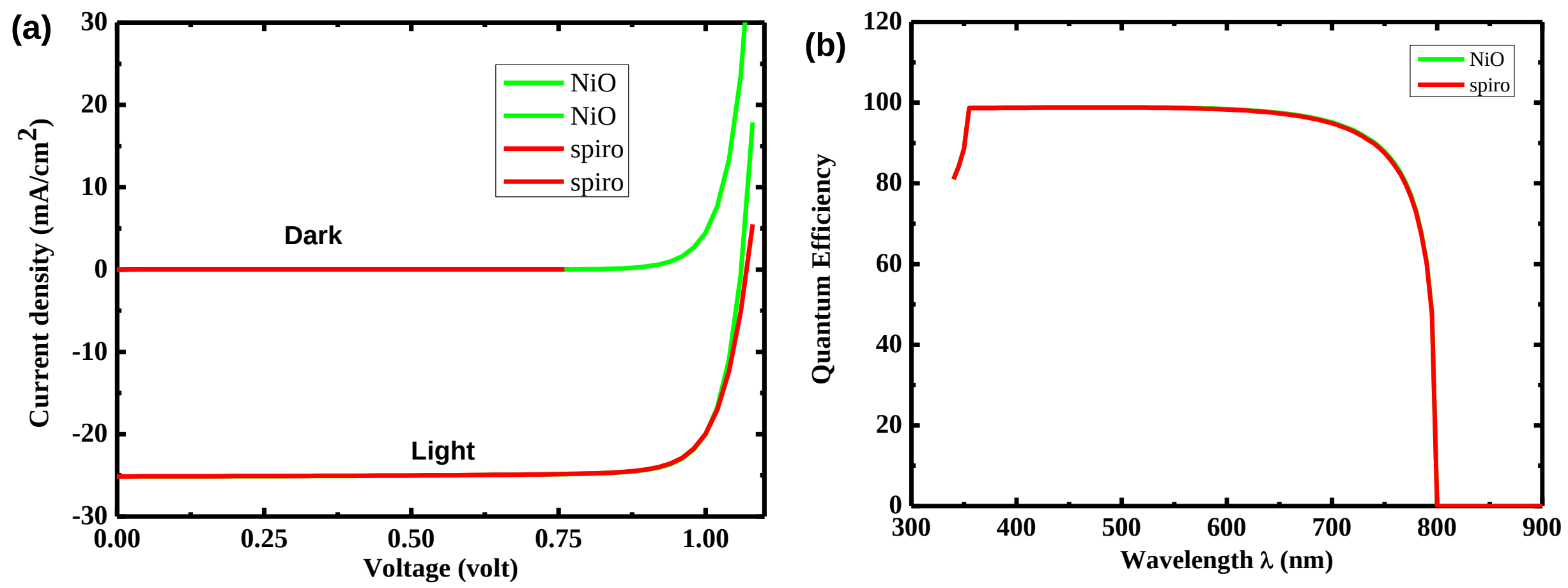

**Figure 6:** (a) J-V characteristic of optimized inverted planner perovskite solar cell under dark and illumination and (b) Quantum efficiency of optimized inverted planner perovskite solar cell under illumination.

### 3.2 Effect of back metal substrate work function and working temperature over efficiency:

In inverted PSC structure, hole transport material is deposited above conducting metal substrate. Till this moment, we considered Ohmic contacts at back metal substrate and hole conductor, using flat band conditions during simulation. In contrast, a suitable work function should be used to find out the suitable metal substrate. Considering this realistic constraints, we varied work function for metal substrate to estimate the optimal work function and thus, a suitable metal substrate which can be considered as the efficient bottom substrate for the deposition of hole transport material. Figure 7(a) shows the efficiency variation against metal substrate work function for both spiro-MeOTAD and NiO hole conductors. We observed that metal with work functions ≥ 5.1 eV and ≥ 5.4 eV are suitable for NiO and spiro-MeOTAD hole conductors. Thus, Au, Co, Ir, Ni, Pd, Pt and Se may be suitable candidates for back metal substrates [43]. Co and Ni are practically suitable for their use as back metal contacts among these metals because of their lower costs and easy availability. However, Ni may be preferred as Ni will have better adhesion and interface properties with NiO hole conductor.

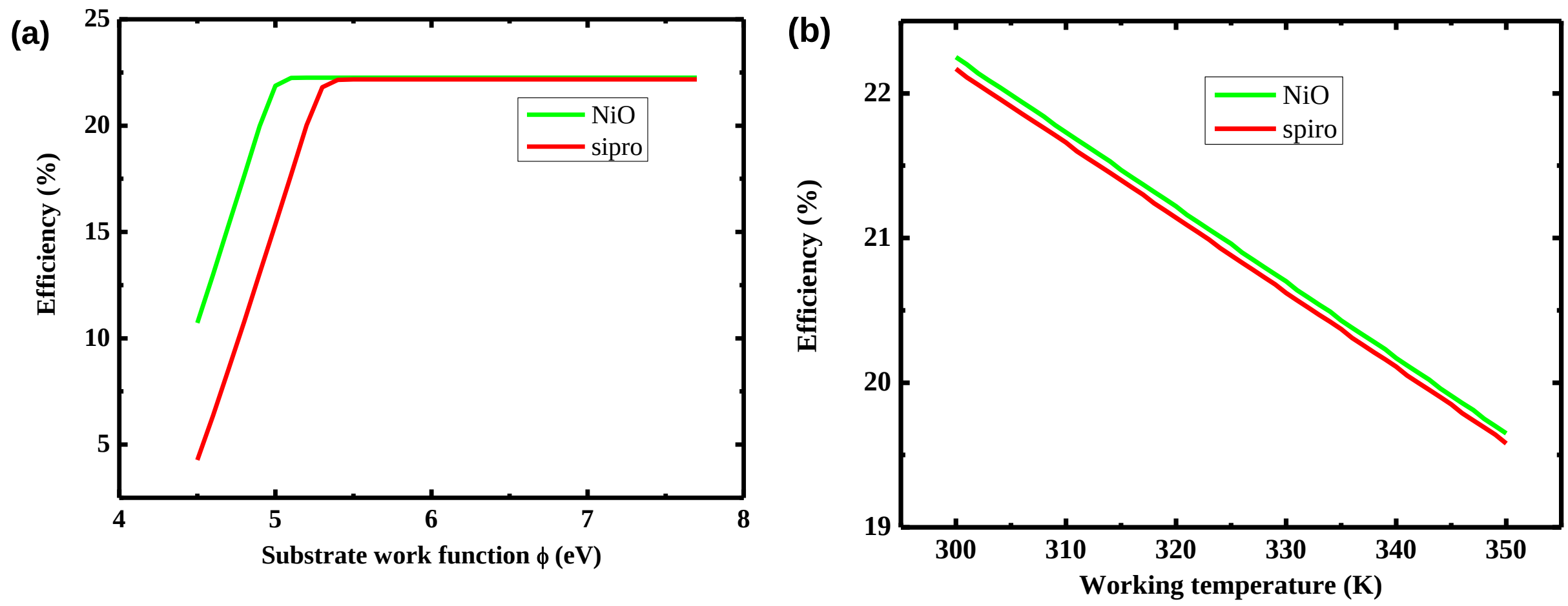

**Figure 7**: (a) Effect of work function of back metal substrate over efficiency of device and (b) effect of working temperature over performance of optimized device.

The working temperature of photovoltaic device can be higher than room temperature (300K), which may finally degrade the performance. Thus, working temperature is varied from 300K to 350K, to observe effect of temperature on photovoltaic response of the optimized device. Results are plotted in Fig 7(b), showing linear decrease with temperature.

### 3.3 Effect of band offset:

Conduction and valence band offsets are important parameters in thin film based solar cells. The band offset study for planner perovskite devices is also performed earlier [21]. The present work explores the band offset effect on inverted PSC structure to see sensitivity of this structure to band offsets. Conduction and valence band offsets are explained schematically in Fig 8. We considered TCO and absorber interface to simulate the effect of conduction band offset and HTM and absorber interface to simulate the valance band offset effects. Conduction and valence band offsets are varied by changing electron affinity of TCO and hole transport material in conjunction with their band gaps. PSCs are evolving and inverted planner device structures are considered for the first time in this study, we are not restricted to FTO as a TCO material and NiO or spiro-MeOTAD as hole conductor materials. Thus, we tried to find set of electron hole conductors for this inverted planner structure with set of band gap and electron affinity.

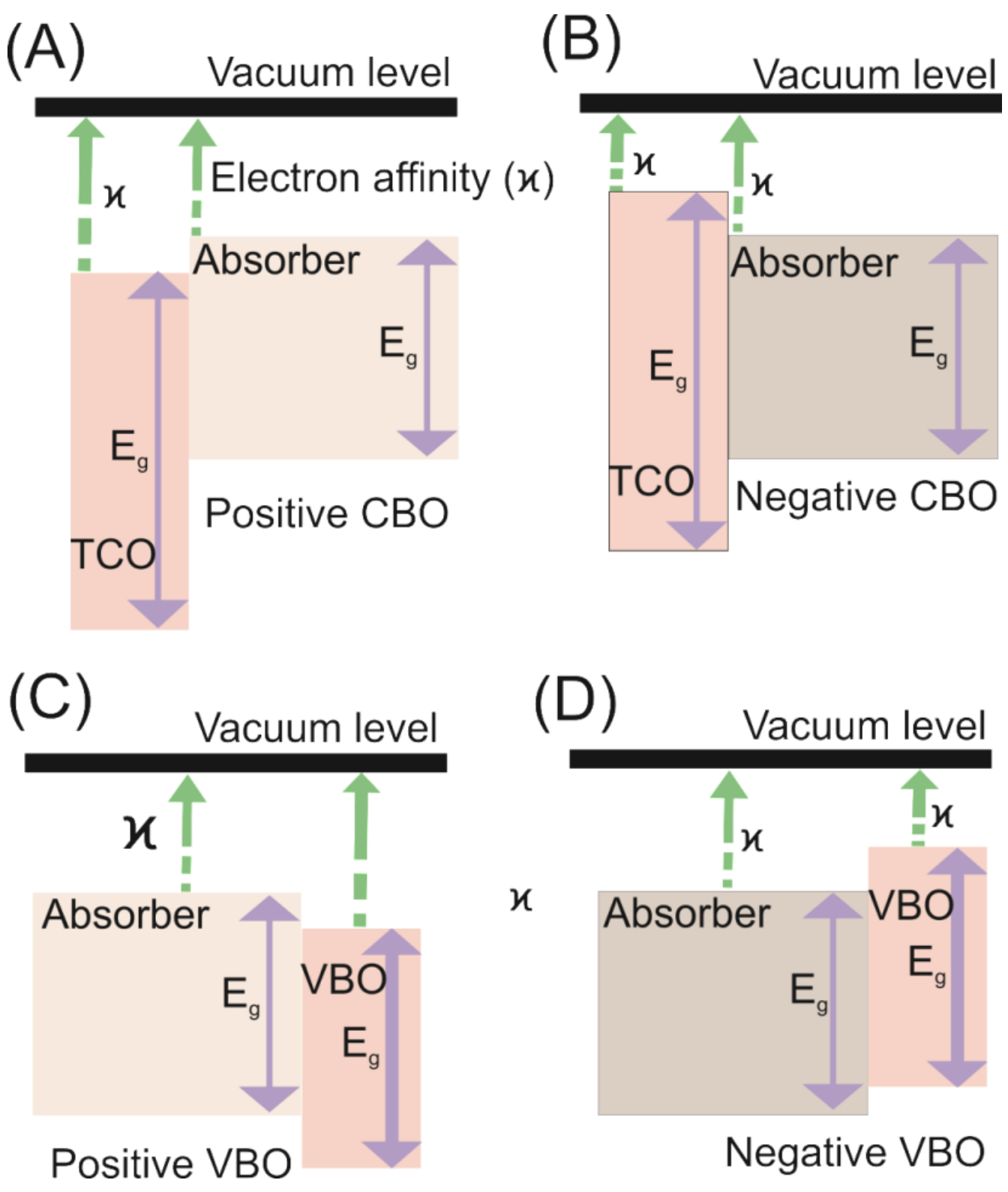

**Figure 8:** (a) Schematic diagram showing positive conduction band offset, (b) schematic diagram showing negative conduction band offset, (c) schematic diagram showing positive conduction band offset and (d) schematic diagram showing negative valence band offset.

Band gap range for TCO and hole conductor is taken as 2.5eV - 3.5eV and 2.5eV - 4.5eV, respectively. Electron affinity values are considered as 3.5eV - 4.5eV and 1.0eV - 2.5eV for TCO and hole conductor, respectively. As the present study is focused on simulation of NiO hole conductor as an equivalent replacement for spiro-MeOTAD in this inverted planner perovskite structure, band offset simulations are carried out for NiO hole material parameter only. Band gap and electron affinity of absorber are not changed during the band offset simulation. Flat band conditions are selected for substrate or left contact and right contact or contact over TCO.

### 3.3.1 Conduction band offset variation:

Conduction band offset defined as difference between electron affinity of TCO and electron affinity of absorber. Fig 8(a) shows schematic diagram for positive conduction band offset and Fig 8(b) shows a schematic diagram for negative conduction band offset. We can see that positive conduction band offset will assist in electron transfer from absorber to TCO while negative conduction band offset will pose barrier in electron transfer from absorber to TCO. To observe effect of conduction band offset, band gap of TCO and electron affinity are varied while band gap and electron affinity of absorber is kept constant with optimized device parameters. Defect density of interface layer is set to be $10^{17}$ $cm^{-2}$. The computed results are plotted in Fig 9 (a). We can see that a dark blue region marked by efficiency of 21.38% shows combination of TCO band gap and electron affinity for good performing device while other region shows existence of barrier at TCO/absorber interface for electron transfer. We can see that there is very wide scope for change of TCO band gap and electron affinity in selected range of parameters for simulation.

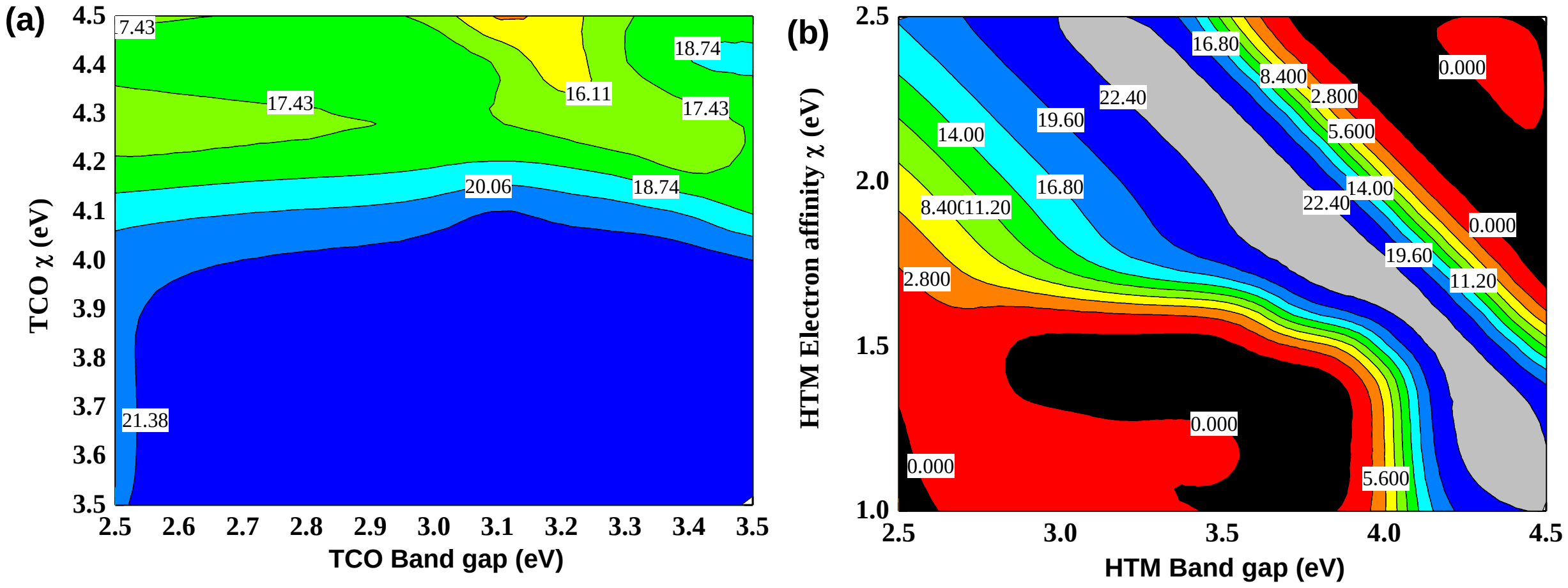

**Figure 9**: (a) Effect over photo-voltaic efficiency of device by change in band gap and electron affinity of TCO and (b) Effect over photo-voltaic efficiency of device by change in band gap and electron affinity of NiO for optimized inverted planner perovskite device.

**3.3.2 Valence band offset variation:**

Valence band offset (VBO) is defined as difference between "electron affinity of absorber plus band gap Eg of absorber" and "electron affinity of HTM plus band gap Eg of HTM". Figs 8 (c) and 8 (d) show schematically positive and negative valence band offsets. Valence band offset depends on the difference between band gap and electron affinity of absorber and band gap and electron affinity of hole transport material. Since, we wanted to simulate valance band offset effect for perovskite absorber, so band gap of absorber in conjunction with electron affinity is kept fixed. To see the effect of VBO on performance, band gap of NiO hole conductor is varied from 2.5 eV to 4.5 eV and electron affinity of hole conductor is varied from 1 eV to 2.5 eV and results are plotted in Fig 9 (b). We could see that for certain combination of band gap and electron affinity of hole conductor, photovoltaic performance is very poor and close to zero (0). These regions are shown by red and black colour in Fig 9 (b). For some combination of band gap and electron affinity values, large band discontinuity is observed and for such cases the simulations could not converge in SCAPS. Such combinations are assigned to zero efficiency. We could observe that silver coloured region, labelled with 22.40% efficiency, represents combination of band gap and electron affinity for hole conductor, providing the optimal performance for inverted planner perovskite devices. We find that there is a narrow region for change in electron affinity and bang gap of hole conductor for valence band offset in contrast to conduction band offset where a large window for band gap and electron affinity is observed.

**4. Conclusion:**

In this work, we proposed a design of inverted planner perovskite solar cell with NiO hole conductor for the first time. The proposed device structure is optimized for different layer thicknesses. The optimized devices are studied against variation in operating temperature, conduction and valance band offset. The increase in temperature led to linear decrease in efficiency of device. Additionally, we observed a very wide range for TCO band gap and electron affinity with conduction band offset provided relatively enhanced device performance while for valance band offset, a narrow range for band gap of hole conductor and electron affinity is observed. These studies will provide a novel inverted PSC device structure and the optimization results will assist in designing a complete PSC device for optimal photovoltaic response.

**Acknowledgement:**

Author Ambesh Dixit highly acknowledges Dr. Marc Burgelman and their team from Gunt University Belgium for providing SCAPS simulator, and Department of Science and Technology, Gov. of India through project # DST/INT/Mexico/P-02/2016 for financial assistance carrying out the present work.

**References:**

[1] and T.M. Akihiro Kojima,† Kenjiro Teshima,‡ Yasuo Shirai, Organometal Halide Perovskites as Visible- Light Sensitizers for Photovoltaic Cells, J Am Chem Soc. 131 (2009) 6050–6051. doi:10.1021/ja809598r.

[2] W.S. Yang, J.H. Noh, N.J. Jeon, Y.C. Kim, S. Ryu, J. Seo, S. Il Seok, High-performance photovoltaic perovskite layers fabricated through intramolecular exchange, Science (80-. ). 348 (2015) 1234–1237. doi:10.1126/science.aaa9272.

[3] H.-S. Kim, C.-R. Lee, J.-H. Im, K.-B. Lee, T. Moehl, A. Marchioro, S.-J. Moon, R. Humphry-Baker, J.-H. Yum, J.E. Moser, M. Grätzel, N.-G. Park, Lead Iodide Perovskite Sensitized All-Solid-State Submicron Thin Film Mesoscopic Solar Cell with Efficiency Exceeding 9%, Sci. Rep. 2 (2012) 1–7. doi:10.1038/srep00591.

[4] M.M. Lee, J. Teuscher, T. Miyasaka, T.N. Murakami, H.J. Snaith, Efficient Hybrid Solar Cells Based on Meso-Superstructured Organometal Halide Perovskites, Science (80-. ). 338 (2012) 643–646.

[5] S.D. Stranks, G.E. Eperon, G. Grancini, C. Menelaou, M.J.P. Alcocer, T. Leijtens, L.M. Herz, A. Petrozza, H.J. Snaith, Electron-Hole Diffusion Lengths Exceeding 1 Micrometer in an Organometal Trihalide Perovskite Absorber, Science. 342 (2014) 341–344. doi:10.1126/science.1243982.

[6] J.M. Ball, M.M. Lee, A. Hey, H.J. Snaith, Low-temperature processed meso-superstructured to thin film perovskite solar cells, Energy Environ. Sci. 6 (2013) 1739–1743. doi:10.1039/c3ee40810h.

[7] M. Liu, M.B. Johnston, H.J. Snaith, Efficient planar heterojunction perovskite solar cells by vapour deposition, Nature. 501 (2013) 395–398. doi:10.1038/nature12509.

[8] Y. Yang, J. You, Z. Hong, Q. Chen, M. Cai, T. Bin Song, C.C. Chen, S. Lu, Y. Liu, H. Zhou, Low-temperature solution-processed perovskite solar cells with high efficiency and flexibility, ACS Nano. 8 (2014) 1674–1680. doi:10.1021/nn406020d.

[9] G.E. Eperon, V.M. Burlakov, P. Docampo, A. Goriely, H.J. Snaith, Morphological control for high performance, solution-processed planar heterojunction perovskite solar cells, Adv. Funct. Mater. 24 (2014) 151–157. doi:10.1002/adfm.201302090.

[10] W.A. Laban, L. Etgar, Depleted hole conductor-free lead halide iodide heterojunction solar cells, Energy Environ. Sci. 6 (2013) 3249. doi:10.1039/c3ee42282h.

[11] A. Mei, X. Li, L. Liu, Z. Ku, T. Liu, Y. Rong, M. Xu, A hole-conductor – free, fully printable mesoscopic perovskite solar cell with high stability, Science (80-. ). 345 (2014) 295–298. doi:10.1126/science.1254763.

[12] D. Liu, J. Yang, T.L. Kelly, Compact layer free perovskite solar cells with 13.5% efficiency, J. Am. Chem. Soc. 136 (2014) 17116–17122. doi:10.1021/ja508758k.

[13] W. Ke, G. Fang, J. Wan, H. Tao, Q. Liu, L. Xiong, P. Qin, J. Wang, H. Lei, G. Yang, M. Qin, X. Zhao, Y. Yan, Efficient hole-blocking layer-free planar halide perovskite thin-film solar cells, Nat. Commun. 6 (2015) 1–7. doi:10.1038/ncomms7700.

[14] E. Zheng, X.F. Wang, J. Song, L. Yan, W. Tian, T. Miyasaka, PbI2-Based Dipping-Controlled Material Conversion for Compact Layer Free Perovskite Solar Cells, ACS Appl. Mater. Interfaces. 7 (2015) 18156–18162. doi:10.1021/acsami.5b05787.

[15] A.A. Zhumekenov, M.I. Saidaminov, M.A. Haque, E. Alarousu, S.P. Sarmah, B. Murali, I. Dursun, X.-H. Miao, A.L. Abdelhady, T. Wu, O.F. Mohammed, O.M. Bakr, Formamidinium Lead Halide Perovskite Crystals with Unprecedented Long Carrier Dynamics and Diffusion Length, ACS Energy Lett. 1 (2016) 32–37. doi:10.1021/acsenergylett.6b00002.

[16] Q. Dong, Y. Fang, Y. Shao, P. Mulligan, J. Qiu, L. Cao, J. Huang, Electron-hole diffusion lengths > 175 um in solution-grown CH 3 NH 3 PbI 3 single crystals, Science (80-. ). 347 (2015) 967–970. doi:10.1126/science.aaa5760.

[17] T. Dullweber, O. Lundberg, J. Malmström, M. Bodegård, L. Stolt, U. Rau, H.W. Schock, J.H. Werner, Back surface band gap gradings in Cu(In,Ga)Se2 solar cells, Thin Solid Films. 387 (2001) 11–13. doi:10.1016/S0040-6090(00)01726-0.

[18] T. Minemoto, J. Julayhi, Buffer-less Cu(In,Ga)Se2 solar cells by band offset control using novel transparent electrode, Curr. Appl. Phys. 13 (2013) 103–106. doi:10.1016/j.cap.2012.06.019.

[19] A. Niemegeers, M. Burgelman, Effects of the Au/CdTe back contact on IV and CV characteristics of Au/CdTe/CdS/TCO solar cells, J. Appl. Phys. 81 (1997) 2881–2886. doi:10.1063/1.363946.

[20] P. Nollet, M. Köntges, M. Burgelman, S. Degrave, R. Reineke-Koch, Indications for presence and importance of interface states in CdTe/CdS solar cells, Thin Solid Films. 431–432 (2003) 414–420. doi:10.1016/S0040-6090(03)00201-3.

[21] T. Minemoto, T. Matsui, H. Takakura, Y. Hamakawa, T. Negami, Y. Hashimoto, T. Uenoyama, M. Kitagawa, Theoretical analysis of the effect of conduction band offset of window/CIS layers on performance of CIS solar cells using device simulation, Sol. Energy Mater. Sol. Cells. 67 (2001) 83–88. doi:10.1016/S0927-0248(00)00266-X.

[22] D. Hironiwa, M. Murata, N. Ashida, Z. Tang, T. Minemoto, Simulation of optimum band-gap grading profile of Cu2ZnSn ( S , Se ) 4 solar cells with different optical and defect properties with different optical and defect properties, Jpn. J. Appl. Phys. 53 (2014) 71201. doi:10.7567.

[23] F. Liu, J. Zhu, J. Wei, Y. Li, M. Lv, S. Yang, B. Zhang, J. Yao, S. Dai, Numerical simulation: Toward the design of high-efficiency planar perovskite solar cells, Appl. Phys. Lett. 104 (2014) 1–5. doi:10.1063/1.4885367.

[24] T. Minemoto, M. Murata, Device modeling of perovskite solar cells based on structural similarity with thin film inorganic semiconductor solar cells, J. Appl. Phys. 116 (2014). doi:10.1063/1.4891982.

[25] T. Minemoto, M. Murata, Impact of work function of back contact of perovskite solar cells without hole transport material analyzed by device simulation, Curr. Appl. Phys. 14 (2014) 1428–1433. doi:10.1016/j.cap.2014.08.002.

[26] T. Minemoto, M. Murata, Theoretical analysis on effect of band offsets in perovskite solar cells, Sol. Energy Mater. Sol. Cells. 133 (2015) 8–14. doi:10.1016/j.solmat.2014.10.036.

[27] K.R. Adhikari, S. Gurung, B.K. Bhattarai, B.M. Soucase, Comparative study on MAPbI3 based solar cells using different electron transporting materials, Phys. Status Solidi Curr. Top. Solid

State Phys. 13 (2016) 13–17. doi:10.1002/pssc.201510078.

[28] L. Huang, X. Sun, C. Li, R. Xu, J. Xu, Y. Du, Y. Wu, J. Ni, H. Cai, J. Li, Z. Hu, J. Zhang, Electron transport layer-free planar perovskite solar cells: Further performance enhancement perspective from device simulation, Sol. Energy Mater. Sol. Cells. 157 (2016) 1038–1047. doi:10.1016/j.solmat.2016.08.025.

[29] M. Hirasawa, T. Ishihara, T. Goto, K. Uchida, N. Miura, Magnetoabsorption of th elowest e xciton in perovskite-type compound (CH3NH3)PbI3, Phys. B Condens. Matter. 201 (1994) 427–430.

[30] M. Burgelman, P. Nollet, S. Degrave, Modelling polycrystalline semiconductor solar cells, Thin Solid Films. 361 (2000) 527–532. doi:10.1016/S0040-6090(99)00825-1.

[31] S. Bansal, P. Aryal, Evaluation of New Materials for Electron and Hole Transport Layers in Perovskite-Based Solar Cells Through SCAPS-1D Simulations, 2016 IEEE 43rd Photovolt. Spec. Conf. (2016) 747–750. doi:10.1109/PVSC.2016.7749702.

[32] M.I. Hossain, F.H. Alharbi, N. Tabet, Copper oxide as inorganic hole transport material for lead halide perovskite based solar cells, Sol. Energy. 120 (2015) 370–380. doi:10.1016/j.solener.2015.07.040.

[33] C.C. Homes, T. Vogt, S.M. Shapiro, S. Wakimoto, A.P. Ramirez, Optical Response of High-Dielectric-Constant Perovskite-Related Oxide, Science (80-. ). 293 (2001) 673–676. doi:10.1126/science.1061655.

[34] P.J. Gielisse, J.N. Plendl, L.C. Mansur, R. Marshall, S.S. Mitra, R. Mykolajewycz, A. Smakula, Infrared properties of NiO and CoO and their mixed crystals, J. Appl. Phys. 36 (1965) 2446–2450. doi:10.1063/1.1714508.

[35] C. Wehrenfennig, G.E. Eperon, M.B. Johnston, H.J. Snaith, L.M. Herz, High charge carrier

mobilities and lifetimes in organolead trihalide perovskites, Adv. Mater. 26 (2014) 1584–1589. doi:10.1002/adma.201305172.

[36] W. Chia-Ching, Y. Cheng-Fu, Investigation of the properties of nanostructured Li-doped NiO films using the modified spray pyrolysis method, Nanoscale Res. Lett. 8 (2013) 1–5. doi:10.1186/1556-276X-8-33.

[37] M. Tyagi, M. Tomar, V. Gupta, Influence of hole mobility on the response characteristics of p-type nickel oxide thin film based glucose biosensor, Anal. Chim. Acta. 726 (2012) 93–101. doi:10.1016/j.aca.2012.03.027.

[38] J.H. Noh, S.H. Im, J.H. Heo, T.N. Mandal, S. Il Seok, Chemical management for colorful, efficient, and stable inorganic-organic hybrid nanostructured solar cells, Nano Lett. 13 (2013) 1764–1769. doi:10.1021/nl400349b.

[39] M.D. Irwin, J.D. Servaites, D.B. Buchholz, B.J. Leever, J. Liu, J.D. Emery, M. Zhang, J.H. Song, M.F. Durstock, A.J. Freeman, M.J. Bedzyk, M.C. Hersam, R.P.H. Chang, M.A. Ratner, T.J. Marks, Structural and electrical functionality of NiO interfacial films in bulk heterojunction organic solar cells, Chem. Mater. 23 (2011) 2218–2226. doi:10.1021/cm200229e.

[40] G. Giorgi, J.I. Fujisawa, H. Segawa, K. Yamashita, Small photocarrier effective masses featuring ambipolar transport in methylammonium lead iodide perovskite: A density functional analysis, J. Phys. Chem. Lett. 4 (2013) 4213–4216. doi:10.1021/jz4023865.

[41] S.J. Fonash, Solar cell device physics, 2nd ed., Academic Press, BURLINGTON, 2010. doi:10.1016/0025-5408(82)90173-8.

[42] T. Leijtens, I.K. Ding, T. Giovenzana, J.T. Bloking, M.D. McGehee, A. Sellinger, Hole transport materials with low glass transition temperatures and high solubility for application in solid-state dye-sensitized solar cells, ACS Nano. 6 (2012) 1455–1462. doi:10.1021/nn204296b.

[43] "Physical Constant of Organic Compound" in CRC Handbook of Chemistry and Physics, 90th Edition (CD-ROM version 2010), D.R. Lide, ed., CRC Press/Taylor and Francis, Boca-Raton, Florida, pp 12-144.